\documentclass[a4paper,10pt]{article}

\usepackage[pdftex]{color,graphicx,hyperref}
\usepackage{amsmath,amsfonts,amssymb}
\usepackage[margin=1in]{geometry}
\usepackage{setspace}
\usepackage{booktabs}
\usepackage[labelfont=bf,labelsep=period]{caption}
\usepackage{subcaption}
\usepackage{scicite}
\usepackage[utf8]{inputenc}
\usepackage[affil-it]{authblk}
\usepackage{comment}
\usepackage{times}

\title{Assortative and preferential attachment lead to core-periphery networks}

\author[1*]{Javier Ureña-Carrion}
\author[2,3]{Fariba Karimi}
\author[4,5,1,6]{Gerardo Iñiguez}
\author[1]{Mikko Kivelä}

\affil[1]{\small{Department of Computer Science, Aalto University School of Science, 00076 Aalto, Finland}}
\affil[2]{\small{Graz University of Technology, 8010 Graz, Austria}}
\affil[3]{\small{Network Inequality group, Complexity Science Hub, 1080 Vienna, Austria}}
\affil[4]{\small{Department of Network and Data Science, Central European University, 1100 Vienna, Austria}}
\affil[5]{\small{Faculty of Information Technology and Communication Sciences, Tampere University, 33720 Tampere, Finland}}
\affil[6]{\small{Centro de Ciencias de la Complejidad, Universidad Nacional Auton\'{o}ma de M\'{e}xico, 04510 Ciudad de México, Mexico}}
\affil[*]{\small{Corresponding author email: javier.urenacarrion@aalto.fi}}

\date{}

\begin{document}

\maketitle

\begin{abstract}
Core-periphery is a key feature of large-scale networks underlying a wide range of social, biological, and transportation phenomena.  Nodes in the core have an influential position in the network, and thus the periphery can be under structural disadvantage if the groups are aligned with external attributes such as gender or economic status.
Despite its prevalence in empirical data, it is unclear whether core-periphery is a consequence of fundamental network evolution processes. While preferential attachment can create degree heterogeneity indistinguishable from core-periphery, it doesn't explain why cores and peripheries are aligned with some external node attribute, i.e. why specific groups of nodes gain dominance and become cores. We show that even small amounts of assortative attachment, e.g., homophily in social networks, can break the symmetric effect of preferential attachment, 
and that the interplay of the two mechanisms leads to one of the groups emerging as a prominent core.
A systematic analysis of the phase space of the proposed model reveals the levels of assortative and preferential attachment necessary for a group to become either core or periphery. 
We find that relative group size is significant, with minority groups typically having a disadvantage on becoming the core for similar assortative attachment levels among groups. We also find that growing networks are less prone to develop core-periphery than dynamically evolving networks, and that these two network evolution mechanisms lead to different types of core-periphery structures. Analyzing five empirical networks, our findings suggest that core nodes are highly assortative, illustrating the potential of our model as a tool for designing and analyzing interventions on evolving networks.

\end{abstract}


\section*{Introduction}

Core-periphery is a ubiquitous property of group-level relationships in networks of social, biological, and infrastructural phenomena \cite{Borgatti2000, Rombach2014, Csermely2013, Gallagher2021}. It involves a notion of meso-scale dominance, where a group of core nodes captures a disproportionate number of connections, whereas the remaining periphery nodes are largely connected to the core and sparsely among themselves. Classical examples of core-periphery structure involve economic and geopolitical networks between countries where a core group dominates trade ~\cite{Kostoska2020trade,Gala2017trade,Gray2012trade}, economic relationships~\cite{Hartmann2020inequality,Bordo2001exchangerate} or soft power~\cite{Snyder1979trade,Lo2011education}, but also encompass networks involving stakeholders or influential individuals~\cite{Villesche2021work}, such as the so-called Old Boys' Club \cite{gamba2001old}. Besides such classical examples, core-periphery structures have been observed in a broad array of empirical phenomena and domains of knowledge, including transportation networks where most routes pass through urban hubs \cite{Verma2016}, social media networks that become centralized depending on the type of information shared \cite{Bastos2018}, and protein interaction networks where structural centrality is related to relevant biological functions \cite{Luo2009}.

Nodes in the core are at a position of power or importance, whereas peripheral nodes have weaker influence. This can become an issue of structural inequality in social, transportation, and economic networks when the core and periphery groups are aligned with some pre-existing, external attribute of nodes. In social, political and cultural contexts, such external grouping might be the gender, language, nationality, religion, or ethnicity of individuals. In transportation and economic networks, groups might be determined by relevant features of the population or country under study. While it is generally understood how core-periphery structures can emerge through network mechanisms such as preferential attachment \cite{moody2003structural,Holme2005, Csermely2013,Verma2016,Lux2015}, these models do not offer an explanation as to when and how do core-periphery structures aligned with external node attributes emerge. Understanding the mechanisms and underlying causes for such emergent structural alignment is an important first step towards policies for reducing the effects of ubiquitous structural inequalities in society \cite{diprete2006cumulative}.

We model emergent core-periphery and its alignment with external attributes in terms of network evolution processes. A common form of topological dominance occurs through heterogeneous degree distributions. The effect is large enough that such distributions alone can explain core-periphery partitions in two-group detection problems, i.e., when determining which nodes belong to the core or the periphery \cite{Kojaku2018}. Heterogeneous degree distributions, in turn, emerge naturally from network evolution processes involving preferential attachment \cite{barabasi1999emergence}. This mechanism is arguably behind many evolving phenomena in nature and society, and while it has been extensively modelled as an explicit driver (where, say, a node is selected with probability proportional to its degree), it also appears implicitly as the consequence of other dynamic processes such as link or vertex copying and triadic closure \cite{Asikainen2020}. Preferential attachment, however, cannot solely explain why certain groups of nodes characterized by relevant node attributes \cite{Borgatti2000} become the core and others the periphery.

To understand how core-periphery structure may arise on networks with a pre-determined notion of groups, we need to include an additional network mechanism based on node attributes. Nodes commonly display a preference for forming links with nodes similar (or dissimilar) to themselves \cite{mcpherson2001birds,Rivera2010}. This mechanism is known as homophily in social networks, or assortative mixing in general \cite{Newman2003}. In sociology, homophily is used as an explanatory mechanism for a wide array of social identities, including group partitions based on, e.g., gender, socioeconomic status, or political affiliation. Homophily has also been linked to emergent network phenomena like community structure \cite{Peixoto2022, Murase2019}, biased perceptions, and biases in the rankings of minority groups \cite{karimi2018homophily, Lee2019}. Assortative attachment is, however, not constrained to sociological contexts and appears in many other networked systems such as biological food networks \cite{Peel2018}, phylogenetic networks \cite{Larremore2013genomic}, and functional brain networks \cite{Lim2019}. Both assortative and preferential attachment encompass broad mechanisms underlying network evolution, which stresses the importance of understanding how they interact in evolving systems.  

We show how the combination of preferential and assortative attachment leads to networks with a strong alignment between predetermined groups and core-periphery structure. Our results show that even small amounts of assortative mixing can break the symmetry of preferential attachment (which favours high-degree nodes irrespective of their group) and lead instead to prominent group-aligned core-periphery structures. These results can be explained by the interaction of these two mechanisms resulting in a cumulative advantage-like cycle \cite{diprete2006cumulative}: preferential attachment favors possible new connections to high-degree nodes, and at the same time, assortativity becomes more effective within the group that becomes the core, even if the other group is also assortative. In cases where both groups are assortative enough, this cumulative process leads to a bifurcation where either of the two groups can end up being the core. Last, while core-periphery can be modelled with pairings of highly assortative and highly disassortative groups, the interaction with preferential attachment both reduces disparities necessary for structural dominance to occur, and cumulatively amplifies the influence of the assortative group.

We use a combination of modelling and empirical data analysis to show the implications of the interplay between preferential and assortative attachment mechanisms on networks. Using a systematic mean-field analysis, we find the amounts of assortative and preferential attachment that lead to core-periphery structures in network models. Our results indicate that the two archetypes of core-periphery structures, hub-and-spoke and layered \cite{Gallagher2021}, are typical outcomes of both rewiring and growing network models. We also examine how relative group sizes may have a large impact on the emergence of core-periphery structures, including which group becomes the dominant one. Finally, by fitting the appropriate types of network models to real-world networks, we compute the size of the effects of interplay between the two mechanisms, and discuss the implications of interventions that could reduce the emergence of core-periphery structures by changing the strengths of preferential attachment in the system.

\section*{Methods}

\begin{figure*}
\centering
\includegraphics{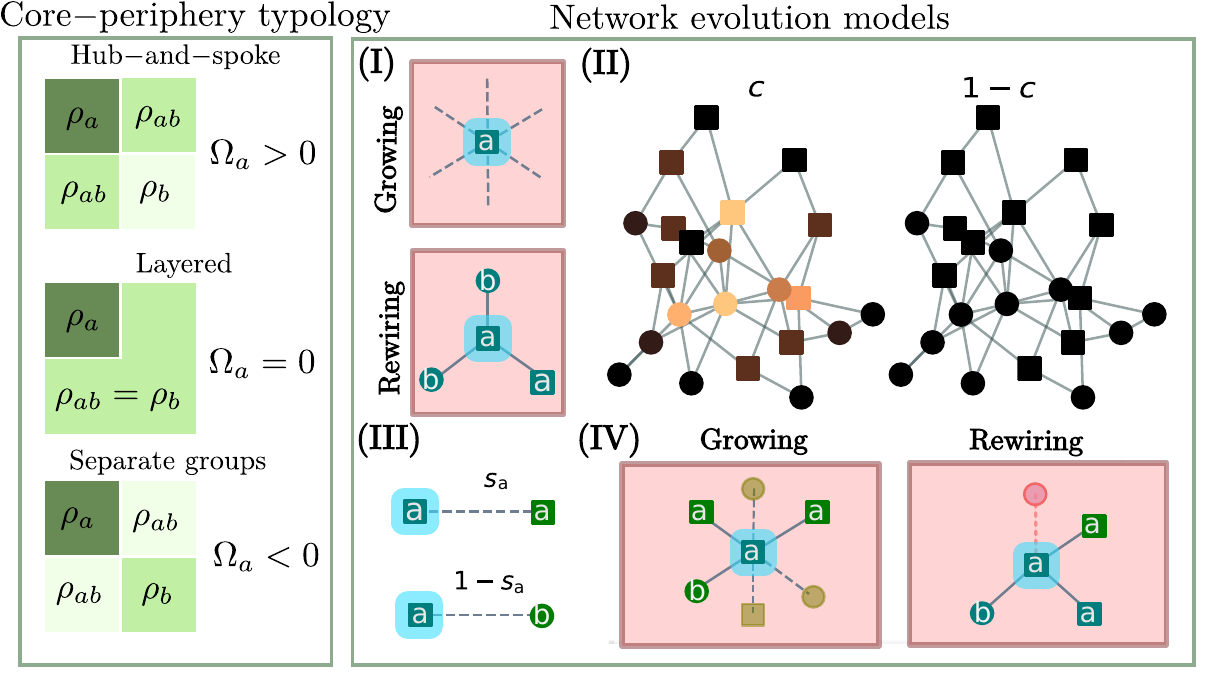}
\caption{\small \textbf{Core-periphery typology and network evolution models.} (\textbf{Left}) Block-model typology of group relations detected by our core-periphery measure $\Omega_a$, with $a$ as core and group densities $\rho_a, \rho_b$, and $\rho_{ab}$. We characterize networks on a spectrum that ranges from hub-and-spoke ($1\geq \Omega_a>0$) and layered networks ($\Omega_a = 0$), both of which are core-periphery relations, to disjoint groups ($-1\leq\Omega_a < 0$). (\textbf{Right}) Main iteration steps of growing and rewiring models used in this work, for two groups (circles and squares). \textit{Step I: Focal node.} In the growing model, a focal node (shaded) arrives in the network, with the node assigned to group $a$ with probability $n_a$. This focal node will find $m$ candidate nodes from the existing network. In the rewiring model, a focal node is selected randomly from the network. \textit{Step II: Candidate selection.} For both models, with probability $c$ a candidate is selected via preferential attachment, i.e. with probability proportional to degree. With probability $1-c$, a candidate is selected uniformly at random. \textit{Step III: Link creation.} If the focal and candidate nodes belong to the same group (e.g., $a$), a link is created with the group's choice homophily probability ($s_a$); otherwise a link is created with probability $1-s_a$. \textit{Step IV: Iteration.} In the growing model, steps II-III are repeated for $m$ candidates, with the focal node possibly rejecting some candidates. In the rewiring model, if the candidate node is accepted, a link is created between the focal node and the candidate. At the same time, the link with a random neighbor of the focal node is removed.}
\label{fig:fig1}
\end{figure*}

\subsection*{Network evolution models}

We develop two models, one for growing and another one for rewiring networks, that combine preferential and assortative attachment and unequal group sizes. 
Our models are inspired by well-known social network mechanisms: a rewiring model mixing homophily and implicit preferential attachment~\cite{Asikainen2020}, and a growing model mixing homophily and explicit preferential attachment~\cite{karimi2018homophily}. Together, our two evolving systems allow us to examine core-periphery structures in a wide variety of systems, as many real-world evolving networks fall into one of these two categories \cite{Rombach2014}. 

Both models share central iteration steps in which a focal node finds a candidate node to connect to (Fig. \ref{fig:fig1} right). In these central steps, a parameter $c$ controls the balance between two candidate selection methods, preferential attachment and random choice. Once a candidate node is selected, the assortativity parameter $s_{ab}$ controls the probability that a focal node from group $a$ connects to a candidate from group $b$. The two models differ in their initial and final steps. For the growing model, the focal node is not connected to the initial network and finds $m$ candidates to connect. In the rewiring model, the focal node is randomly selected, and if a link is created, then a connection to a random neighbor is deleted from the focal node. 

Instead of a matrix $\{ s_{ab} \}$ of four independent assortativity parameters, we only use two free parameters ($s_{aa}=s_a$ and $s_{bb}=s_b$), which also determine the complementary probabilities of choosing a candidate node from the other group ($s_{ab}=1-s_a$ and $s_{ba}=1-s_b$). This approach reflects the actual acceptance rate for the focal node's group, in the sense that group $a$ is assortative if $s_a>0.5$ and disassortative otherwise. 
To see this, given $s_{aa}$ and $s_{ab}$, we show that there exists a reparameterization constant $k$ that defines new variables $\hat{s}_{aa}=ks_{aa}$ and $\hat{s}_{ab}=ks_{ab}$, such that $\hat{s}_{ab}=1-\hat{s}_{aa}$. Setting $k=\frac{1}{s_{aa}+s_{ab}}$ meets that requirement, and so $\hat{s}_a=\hat{s}_{aa}$ is our chosen free parameter for group $a$. This reparameterization, valid for $0\leq \hat{s}_{a} \leq1$,  impacts the evolution rate of the model, but not its fixed points.

\subsection*{Characterizing core-periphery structures}

Core-peripheries were first introduced in social network analysis by Borgatti and Everett \cite{Borgatti2000}. Despite the intuitive notion of group dominance, precise definitions and measurements of core-periphery have been the subject of debate \cite{Csermely2013, Gallagher2021}. Borgatti and Everett defined ``ideal'' group dominance as a fully-connected core, a spectrum of inter-group mixing, and a periphery disconnected from itself, and measured core-periphery as the correlation between the empirical adjacency matrix and an ideal matrix of the same size. Many authors have since then proposed various definitions of core-periphery with different underlying assumptions or for specific domains \cite{Borgatti2000, Csermely2013, Rombach2014, Gallagher2021}. Some of the most common formulations focus on group densities \cite{Zhang2014, Kojaku2017, Gallagher2021}, but also include $k$-cores \cite{Holme2005}, notions of centrality \cite{Csermely2013}, and network capacity \cite{DaSilva2008}. More recently, Gallagher \textit{et al.}~\cite{Gallagher2021} have proposed a typology of core-periphery with two main classes based on group densities: \textit{hub-and-spoke} networks as a generalization of Borgatti and Everett's ``ideal'' structures, and \textit{layered} networks, where the periphery is as likely to connect to itself as to the core. 

Given the various approaches to characterizing core-peripheries in the literature, we choose two metrics that capture different aspects of network structure, both requiring a prior definition of which group is a potential core. For a core group $g\in\{a,b\}$ where $a, b$ are the two groups in the network, we introduce $\Omega_g$ as a measure of the difference between densities of core and periphery, which leads to the classification of hub-and-spoke and layered networks as in \cite{Gallagher2021} (Fig 1, left). We focus on the group with the largest density, which we refer to as \textit{density-dominant}. For example, group $a$ is density-dominant if $\rho_a > \max\{\rho_{ab}, \rho_b\}$, with $\rho_g$ the fraction of links in group $g$, and $\rho_{ab}$ the fraction of inter-group links. Then $\Omega_a$ takes the form 
\begin{equation}
\Omega_a = \frac{\rho_{ab}-\rho_b}{\rho_{ab}+\rho_b}\,,
\end{equation}
which is $1$ for the ideal hub-and-spoke structure, $0$ for an layered network, and $-1$ for two disjoint groups (Fig. \ref{fig:fig1} left). 

As a second measure of core-periphery, we consider Borgatti and Everett's notion of an ``ideal'' core-periphery network via the quantity $r_g$, a continuous re-parametrization of the correlation between discrete adjacency matrices \cite{Borgatti2000}. Using a continuous form allows us to use it on the network statistics of our mean-field analysis. 

All of our theoretical and empirical analyses are based on the group mixing matrix $P$, defined as $P_{ab}=\frac{L_{ab}}{L}$, where $L_{ab}$ is the number of undirected links between groups $a$ and $b$, and $L$ is the total number of links. Both measures of core-periphery can be written in terms of $P$, the total density $\rho=\frac{2L}{N(N-1)}$ for $N$ nodes, and the minority group size $n_a$. We can derive $\Omega_g$ via the approximation of local densities (derivation in Supplementary Information [SI])
\begin{equation}
\rho_{a} \approx P_{aa}\frac{\rho}{n_a^2}.
\end{equation}

For $r_g$, we use a continuous form of the adjacency matrix. Let $n_a$ and $n_b=1-n_a$ be the relative number of nodes in each group, and $0<\alpha\leq1$ a continuous parameter that regulates the distribution of inter-group links. The the continuous form of the Borgatti and Everett measure when group $a$ is the core is (derivation in SI) 
\begin{equation}
\label{eq:becont}
r_a(\alpha) = \frac{\sqrt{\rho}\left(P_{aa} - n_a^2 + \alpha P_{ab}  - 2\alpha n_an_b\right)} {\sqrt{(1-\rho)(n_a^2 + 2\alpha n_an_b)(1-n_a^2 -2\alpha n_an_b)}}.
\end{equation}

We find $r_g$ by maximizing over all possible $\alpha$ values, so that $r_g=\max_{\alpha}{r_g(\alpha)}$. The measure $r_g$ is largely dependent on statistics from the core (such as $P_{aa}$ and $n_a$ in equation \ref{eq:becont}) but not from the periphery. 
We interpret this measure as correlation to an ideal; however, we frame it in terms of ``core quality", as it places particular focus on links within and, to a lesser extent, from the core.
In other words, $r_g$ is larger when the density of the core explains a larger fraction of the overall density of the network. 

Notably both discrete and continuous $r_g$ measures are dependent on overall network density, which 
implies that comparing $r_g$ values is only possible for networks of the same density. We use fixed $\rho=0.1$ in all theoretical analyses to render them comparable, and discuss additional limitations of $r_g$ in the SI. 

\section*{Results}

\subsection*{Mean-field equations}

We begin by a systematic analysis of our models, with a focus on examining the fixed points that describe the final state of the network after the mechanisms have had sufficient time to operate. 
We model the evolution of group interactions in both of our network models using a combination of approximate mean-field equations (MFE) and simulations. 
The former capture the group-level dynamics of our network models by representing the intra- and inter-group link distributions in a continuous fashion, whereas the latter serve as a validation for the MFEs.
In our mean-field methodology, we first find the fixed points associated with group mixing and then characterize their level of core-periphery across parameter space. We compare these results with simulated networks of 5,000 nodes for 100,000 time steps from three different initial conditions. 
    
We track the evolution of MFEs for the group mixing matrix $P$. Since the elements of $P$ satisfy the linear constraint $P_{aa}+P_{ab}+P_{bb}=1$, it suffices to characterize the evolution of the intra-link distributions via $\frac{d P_{aa}}{d t}$ and $\frac{d P_{bb}}{d t}$ for each of our models. 
Both growing and rewiring models share a candidate-selection process, which constitutes the main building block of our algorithms. 
This is mirrored in the MFEs by considering a matrix $M$, where $M_{ab}$ is the probability of creating a link between groups $a$ and $b$. With probability $c$ we find a candidate node from $b$ via preferential attachment, and with probability $1-c$ we find it via random choice; this candidate is then accepted with probability $s_{ab}$, so  
\begin{equation}
M_{ab} = \left[c\left(\frac{1}{2}P_{ab} + P_{bb}\right) + (1-c)n_b \right]s_{ab},
\label{eq:m_mat}
\end{equation}    
    
which is derived from a formulation in terms of total degrees $K_a$ and $K_b$, that is, the sum of degrees of all nodes in groups $a$ and $b$, respectively (see SI). 
The matrix $M$ is not symmetric, as the probability of creating a link from $a$ to $b$ means selecting nodes from $b$ only. 

The MFEs for each model are a function of the matrix of link-creation probabilities $M$, relative group sizes $n_a$ and $n_b$, and the group mixing matrix $P$.
Starting with the rewiring model, we track $\frac{d P_{aa}}{d t}$ as the difference between added links $P_{aa}^+$ and deleted links $P_{aa}^-$, so that $\frac{d P_{aa}}{d t} = P_{aa}^+ - P_{aa}^-$. We define $T_{ab}$ as the probability that, starting from group $a$, we follow a link and end up in group $b$. In terms of the matrix $P$, $T_{aa} = \frac{2*P_{aa}}{2*P_{aa} + P_{ab}}$. Using this notation, the number of added links depends on the probability of selecting $a$ and then creating a link to $a$, $P_{aa}^+ = n_a M_{aa}$. Then, $P_{aa}$ will decrease under the scenario that: (1) a link is created starting form $a$ to any group ($M_{aa}+M_{ab}$); and (2) a random neighbor in group $a$ is deleted (which happens with probability $T_{aa}$). Therefore, $P_{aa}^-= n_a T_{aa}(M_{aa} + M_{ab})$.
    
By definition, the growing model only adds links, so $P_{aa}$ only decreases when new links are connected to group $b$. Thus, we write the growing MFEs in terms of the number of edges $L$,
\begin{equation}
\frac{d P_{aa}}{d t} = \frac{d}{d t}\left(\frac{L_{aa}}{L}\right)  = \frac{1}{L}\left( \frac{d L_{aa}}{d t} - \frac{d L}{d t}P_{aa} \right).
\label{eq:mfe_grow}
\end{equation}
We model $\frac{d L_{aa}}{d t} = mn_aM_{aa}$ with $m$ the number of candidate neighbors for each new arrival, while $\frac{d L}{d t} = \sum_{ij}\frac{d L_{ij}}{d t}$. Although Eq. (\ref{eq:mfe_grow}) depends on the total number of links $L$, such dependency becomes negligible at the fixed points for the matrix $P$ (for derivation see SI).

\subsection*{Assortative and preferential attachment lead to core-periphery}

Our main results can be summarized in three key points. First, core-periphery is an emergent structure of both growing and rewiring models. Second, these models yield qualitatively different core-periphery structures: the growing model is more likely to produce a diffuse core (low but positive $r_g$) and a layered network ($\Omega_g \approx 0$). The rewiring model leads to more acute structural changes, since it is more likely to produce a sharp hub-and-spoke network with a clear core (large $r_g$ and $\Omega_g > 0$). Last, 
despite the qualitative differences between models, both can produce a wide spectrum of core-peripheries across the parameter space, and capture other configurations like disjoint communities and fully mixed groups.

\begin{figure}
\centering
\includegraphics{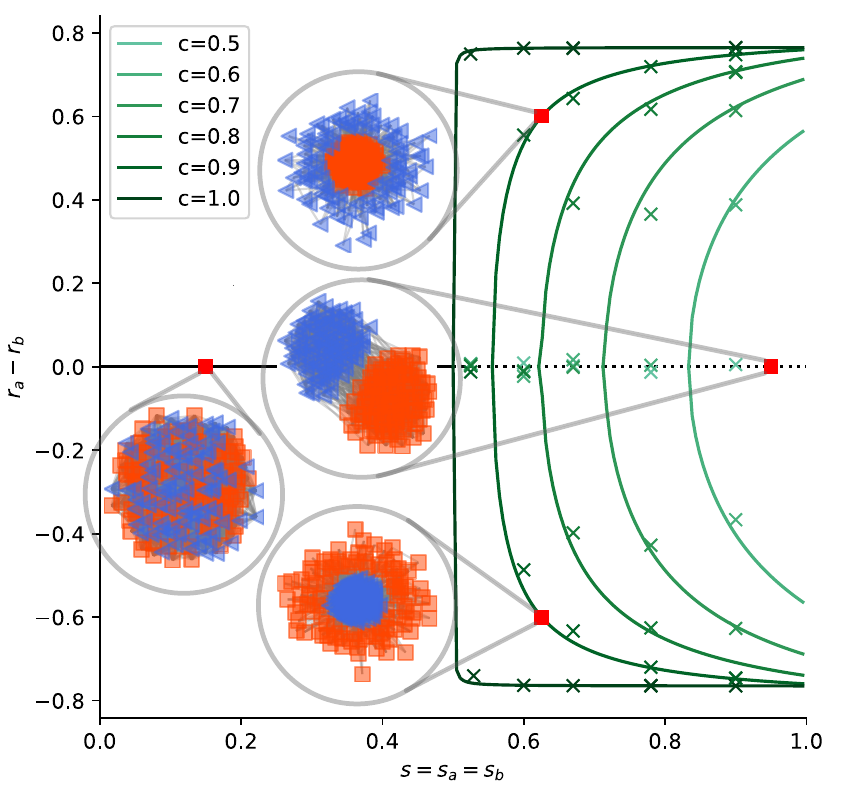}
\caption{\label{fig:fig2_amp}\small \textbf{Homophily and preferential attachment lead to core-periphery.} Core-periphery emergence in terms of differences in correlation to the ideal ($r_a-r_b$) in the rewiring model as a function of assortativity ($s=s_a=s_b$) and preferential attachment $c$, with equal group sizes $n_a=0.5$ and fitted simulation values (crosses). Large homophily and preferential attachment lead to a bifurcation where either group can become a core. The measure $r_a-r_b$ helps visualize this bifurcation, since it is positive (negative) when $a$ ($b$) is a core. The dotted line represents unstable and stable solutions for different $c$ values: if $c$ has bifurcated for given $s$, the dotted line is an unstable solution; otherwise the solution is stable. 
Insets: visual representations of networks with 
$N=200$ nodes and $L=1000$ links at four stable points.}
\end{figure}

Core-periphery emerges from the interplay of assortative and preferential attachment (see fixed points of our rewiring model in Fig. \ref{fig:fig2_amp}). In a stylized scenario with a single assortative attachment parameter and equal group sizes, any amount of assortativity ($s>0.5$) will create strong core-peripheries in the presence of full preferential attachment ($c=1$). When some preferential attachment is replaced with uniformly random selection ($c<1$), core-peripheries still emerge, but a larger amount of assortative attachment $s$ is required to reach the transition point. Note that in this scenario both groups can become cores, since the MFEs have two stable fixed points corresponding to the case where each group is the core ($r_a-r_b$ either positive or negative). This happens because no group has clear assortative or size dominance, resulting in two symmetric fixed points dependent on initial conditions or small stochastic fluctuations. In the rewiring model, such dependence occurs when both groups are assortative, including when one group is smaller than the other (for dependence on initial conditions see SI). The lack of core-periphery ($r_a - r_b = 0$) can imply vastly different network configurations: if both groups are disassortative the resulting network is mostly mixed, but in an assortative scenario the network will consist of mostly disjoint groups.

\begin{figure*}
\centering
\includegraphics[width=.75\textwidth]{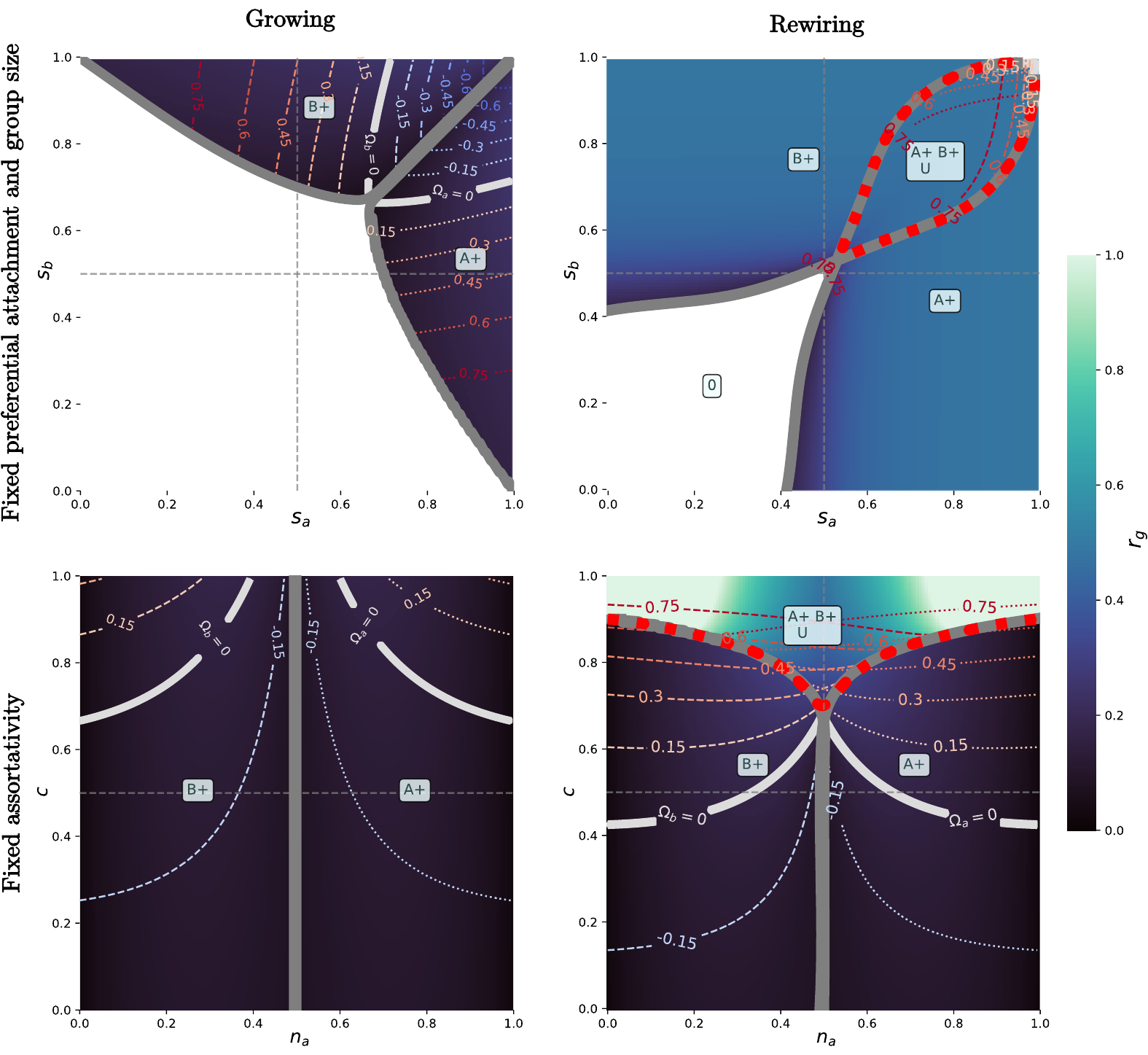}   
\caption{\small \textbf{Fixed-point phase space analysis.} Two measures of core-periphery for (left) growing and (right) rewiring models. Bold grey lines enclose areas where a group $g$ is density dominant (e.g., for $A+$ we have $\rho_a > \max\{\rho_{ab}, \rho_b\}$). Colors depict $r_g$ and dotted lines contours of $\Omega_g$, with emphasis on $\Omega_g=0$ (a full onion network). In the rewiring model, bold dotted red lines enclose areas with three fixed points, where both groups can become a core ($A+$, $B+$), and an additional fixed point is unstable ($U$).
Strict hub-and-spoke core periphery occurs when $\Omega_g>0$. (\textbf{Top}) Models with fixed high preferential attachment ($c=0.95$) and equal group sizes ($n_a=0.5$), varying homophily. (\textbf{Bottom}) Fixed associativity ($s_a=s_b=0.75$) varying preferential attachment and minority size.}
\label{fig:fig3_ps}
\end{figure*}

We explore the phase space of parameters in both models (Fig. \ref{fig:fig3_ps}). The growing model has single fixed points over the whole phase space, producing diffuse cores when a group is density-dominant; i.e., in the regions where a group's density is larger than other local densities, there is little correlation to the ideal core-periphery. For fixed assortativity, higher values of preferential attachment largely result in layered networks ($\Omega_g\approx 0$). In contrast, the rewiring model generates cores with a higher correlation to the ideal hub-and-spoke structure (large $r_g$), as well as increasingly less dense peripheries ($\Omega_g$ grows along with $c$).

When comparing both models over the same phase space, rewiring has a symmetry-breaking effect on the baseline growth model. The network behaves as sharp hub-and-spoke (large $\Omega_g$) as long as one of the groups is assortative, and even slight disassortativity ($s_g>0.4$) can lead to a group becoming the core if the other group is more disassortative. 
When both groups are assortative in the rewiring dynamics, either can become the core. The symmetry-breaking effect also occurs when both groups are equally homophilous (lower panel in Fig. \ref{fig:fig3_ps}). Whereas for the growing model the density-dominant group is usually the larger group, in the rewiring model increasing preferential attachment results in either group becoming a core. Although our models show that larger groups can more easily become a core when assortativity parameters are equal, preferential attachment can boost an assortative minority group ($n_g<0.5$) to become a core (see additional regions of the phase space in SI).
These differences between growing and rewiring models are likely explained by link rigidity in the growing model, where a given link remains static once it has been added to the network (conversely, links in the rewiring dynamics are constantly shuffled).
    
Note that the assortative attachment parameters alone can explain core-periphery relations in both growing and rewiring scenarios (see SI), e.g., when one group is assortative and the other disassortative. However, preferential attachment changes these baseline relations and induces sharper core-periphery configurations, particularly in terms of $\Omega_g$.

\subsection*{Likelihood functions for evolution models} 

We use a temporal maximum likelihood (ML) approach to fit both of our models to real-world data. We use ML estimates for parameters $\theta = (s_a, s_b, c)$, and assume $n_a$ to be directly observed from the data. 
We model the temporal dynamics as a Markovian process: the probability of observing $G_t$, a network at time $t$, solely depends on the network at the previous time step $P(G_t|G_{t-1}, \theta, i_t)$, where $i_t$ is the group of the focal node at time $t$, and is given by the data. For $T$ time steps, the log-likelihood function takes the form 
\begin{equation}
l(\theta) = \sum_{t=1}^T \log P(G_t|G_{t-1}, \theta, i_t).
\end{equation}

The step-wise likelihoods $P(G_t|G_{t-1}, \theta, i_t)$ are based on the multinomial distribution, although they differ for the growing and rewiring models. The multinomial distribution captures how $m$ independent trials are divided between $\hat{k}$ categories. As an example, using the $M$-matrix notation, a focal node from group $b$ might choose a rewiring candidate ($m=1$) from $\hat{k}=2$ categories $a$ or $b$ with probabilities $M_{ba}$ and $M_{bb}$, respectively. Although a plausible likelihood function, this form is mathematically degenerate for estimating $c$, so we use a degree-based approach. We model the probability $p_k^{ab}$  that a focal node from group $a$ creates a link to a node of degree $k$ in group $b$. Considering that $n^b_k$ is the number of nodes in group $b$ of degree $k$, and obviating the notation for temporal data $t$, this probability is
\begin{equation}
p^{ab}_k = \left(c \frac{n^b_kk}{\sum_j(n^a_jj + n^b_jj)} + (1-c)  \frac{n^b_k}{\sum_j(n^a_j + n^b_j)}\right)s_{ab}.
\end{equation} 

Denoting by $x_k^{ia}$ the observed number of links created from a focal node in group $i=i_t$ to a node of degree $k$ in group $b$, and generalizing $m$ trials at each time-step, we have that the likelihoods at time $t$ for the rewiring model are
\begin{eqnarray}
&P(G_t|G_{t-1}, \theta, i_t)=\frac{m!}{\prod_{k}x^{ia}_k(t)!x^{ib}_k(t)!}\label{eq:stepllik}\\
& \times\prod_k p^{ia}_k(t)^{x^{ia}_k(t)}p^{ib}_k(t)^{x^{ib}_k(t)}. \nonumber
\end{eqnarray}
The likelihood function for the rewiring model follows Eq.~(\ref{eq:stepllik}) when $m=1$. The growing model, however, includes several candidate connections ($m\geq1$) and an unobserved number of rejected candidates (denoted $x^{in}$ with probability $p^{in}$). We obtain the marginal probability distribution by integrating over the unobserved number of rejected candidates, thus yielding two different expressions for the likelihood functions of our models (for details on these functions see SI). The functions do not have simple analytic solutions, and thus we use a numerical optimization approach with a BFGS algorithm \cite{Fletcher1987}. We validate our procedure by reconstructing the parameters from simulated data (details available on SI). 

The focus on parameter inference means that the non-parameterized mechanistic components of our model are not modelled. These components include the randomized selection of the focal node (assumed to occur with probability $n_a$) and the link deletion step. In practice, the focal node is determined by the data itself, and link deletion in the rewiring model occurs when a link's lifespan is longer than a sliding window $\Delta t$, and not when a focal node creates a new connection. These deviations could effect differences between observed and predicted group mixing matrices $P$; however, we found this approach to be the least invasive solution for disparities between our evolution model and the empirical data (details of data pre-processing to fit our models are available on the Appendix).

\subsection*{Evolution mechanisms in empirical networks with core-periphery}
    
We showcase an empirical association between preferential attachment, assortativity and core-periphery, and also illustrate the potential of our models in predictive settings by fitting five network datasets using ML estimates. We select datasets with the criteria that they are temporal in nature and have at least two predetermined groups of nodes. These datasets cover different fields and span large observation periods, including Twitter discussions about climate politics collected in Finland for 11 months, with the minority group composed of political actors (\textit{Twitter}); the evolution of gendered dynamics in boards of directors over 9 years (\textit{Boards}); two citation networks with groups determined both by geography (\textit{Cit-geo}) and sub-fields within physics (\textit{APS}); and airport use networks in the United States (\textit{Airport}). When discussing social networks, we refer to assortative attachment as choice homophily \cite{Asikainen2020}. For each dataset, we fit the model that is more likely to capture the evolution mechanisms, so that our two citation datasets are growing networks, while \textit{Twitter}, \textit{Boards} and \textit{Airport} are rewiring (even if these datasets also include node growth, we focus on link rewiring as the more prominent mechanism). All datasets allow for the construction of various networks based on metadata or different temporal and geographical aggregations. We systematically filter for networks that (i) display some degree of core-periphery and (ii) show evidence of preferential attachment (see Appendix for details on each dataset).

\begin{figure*}
\centering
\includegraphics{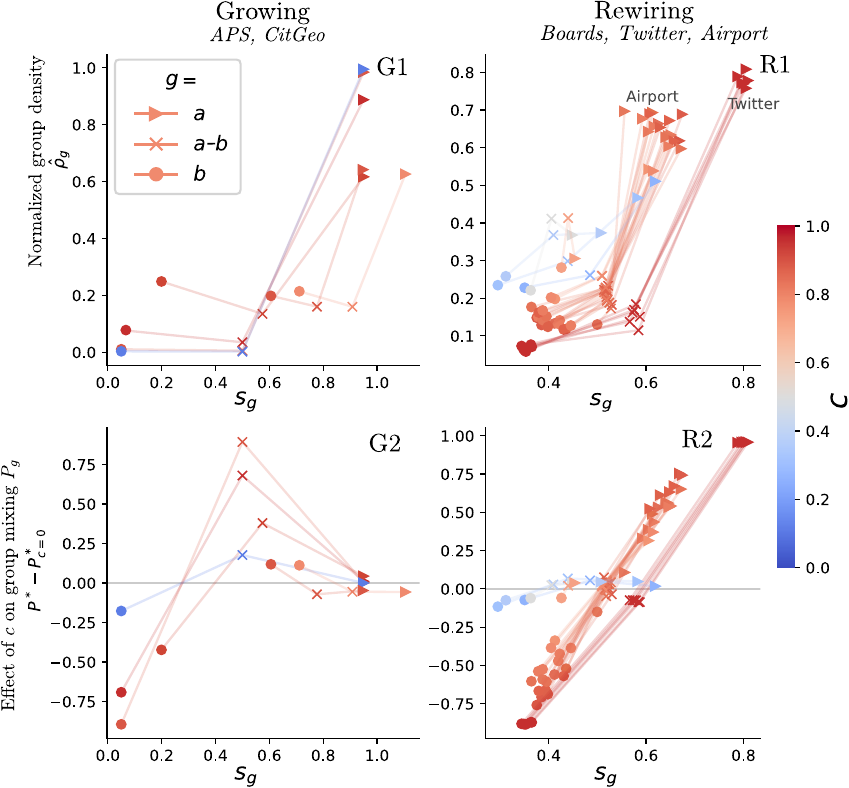}
\caption{\label{fig:fig4_emp}\small \textbf{Empirical association between estimated parameters and group densities in core-peripheries, and effect of preferential attachment given parameters.} For each plot, each line represents a network, with the x-axes depicting the estimated assortative mixing parameters ($s_a$ by triangles, $s_b$ by dots) for both groups under (left) growing and (right) rewiring models. For the visualization of inter-group density $\hat{\rho}_{s_{a-b}}$, we define $s_{a-b}=\frac{s_a+s_b}{2}$. The (\textbf{Top}) row depicts the association to observed densities at the end of the observation period, and the (\textbf{Bottom}) row depicts the estimated effect of removing preferential attachment.
Here, where positive (negative) values imply gained (lost) links in a group. 
We observe that minority nodes ($g=a$), which are the core in most of our datasets, tend to be highly assortative, while growing (G1) and rewiring (R1) models follow the typologies of layered and hub-and-spoke structures based on the order of group densities, with an average $\Omega_g=.224$ for rewiring networks and $\Omega_g=-0.04$ for growing networks. This is likely explained by the effect of preferential attachment that increases inter-group mixing in the growing model (G2) and core-mixing in the rewiring model (R2). 
We normalize group densities to render datasets visually comparable as empirical densities vary widely. The normalization $\hat{\rho}_a=\frac{\rho_a}{\sum_{g}\rho_g}$ then affects the range but not the order of group densities.  
To see the effect of preferential attachment we
compare with a homophilous-mixing baseline. $P^*-P^*_{c=0}$ depicts the difference in fixed points between the estimated full model ($c, s_a, s_b, n_a$) and the model for $c=0$, where positive (negative) values imply a loss (gain) of links on the entries of the matrix $P$. For the \textit{Airport} and \textit{Twitter} datasets, estimated parameters cluster together.
}
\end{figure*}

Our data analysis confirms several key results regarding core-periphery in networks. 
Fig. \ref{fig:fig4_emp} depicts both the empirical association and theoretical predictions of our models, where each group's density is related to their estimated assortativity.
We find that the predefined groups in the data have an aligned core-periphery structure. In such alignments, the core groups are more assortative than peripheries ($s_a > s_b$). 
We corroborate that the evolution mechanisms follow their more common typologies, i.e. rewiring networks are hub-and-spoke, whereas growing networks fall between layered and slightly disjoint structures. This likely occurs due to the effect of preferential attachment, which largely increases inter-group links for the growing model, and intra-core links for the rewiring models. 

Our model predicts that the preferential attachment mechanism strongly amplifies core-periphery structures that would otherwise be weak if solely explained by different assortative preferences within core and periphery nodes. In order to quantify this, we examine the effect of preferential attachment via the the group mixing matrix $P$, defined as the fraction of links that fall within/between groups, $P_{ab}=\frac{L_{ab}}{L_{aa}+L_{ab}+L_{bb}}$. We compare $P^*$ and $P^*_{c=0}$, i.e. the MFE fixed points using the estimated parameters $P^*$ against the fixed points with no preferential attachment ($c=0$). 
We find that in the growing model preferential attachment increases the number of inter-group links at the expense of the periphery when the periphery is disassortative, while slightly increasing the number of within-periphery links when it is assortative. Notably, for the rewiring model preferential attachment vastly increases the fraction of links within the core at the expense of links within the periphery. 
    
The \textit{Twitter} dataset of discussions of climate policy is an interesting example of a network with a homophilous core and a heterophilous periphery. The peripheral majority group (composed of non-politician Twitter users) displays heterophilous behaviour. This could be related to the fact that the minority group (politicians), are clear stakeholders in public policy, which implies a form of influence.
Indeed, such public positions might already be a form of preferential attachment in the sense that people retweet politicians because of their importance outside of Twitter. From our model's perspective, this is captured as a heterophilous choice to retweet; and indeed such heterophilous-homophilous pairing could already explain some forms of core-periphery.
However, modeling preferential attachment in terms of node degrees allows us to see a distinct effect: retweeting popular people means that even more people are exposed to them in a way that cumulatively amplifies their messages. In the end, our models strongly suggest that these compound effects yield structural group dominance in the overall network where the peripheral nodes are more likely to be connected to politicians than to non-politicians, while politicians form a tight net. This result is in line with Ref. \cite{Bastos2018}, which
reports that discourse networks on Twitter shift to core-peripheries when sharing specialized information from government agents (forming the core), but become decentralized in generic discussions.
We highlight that this analysis does not account for a polarized discourse. However, one could hypothesize that different political parties have central roles within their own political spheres.

Our empirical analysis is constrained by some limitations that hinder our capacity to predict the mixing patterns in $P$ for some datasets. Real-world data might deviate from some of our models' assumptions. For example, nodes might not be uniformly active in link creation but display more heterogeneous behaviour. Nodes might also follow additional mixing patterns, e.g., other assortativity groupings based on political positions on Twitter. In the case of the rewiring model, link deletion might not necessarily follow from link creation. Despite these shortcomings, our model can provide further insights in datasets where these constraints are not as strong. In the following section we examine an application where parameters are not static, showing how parameter changes affect mixing patterns and core-periphery. 

\subsection*{An application: Effect of interventions in social networks}

\begin{figure*}
\centering
\includegraphics{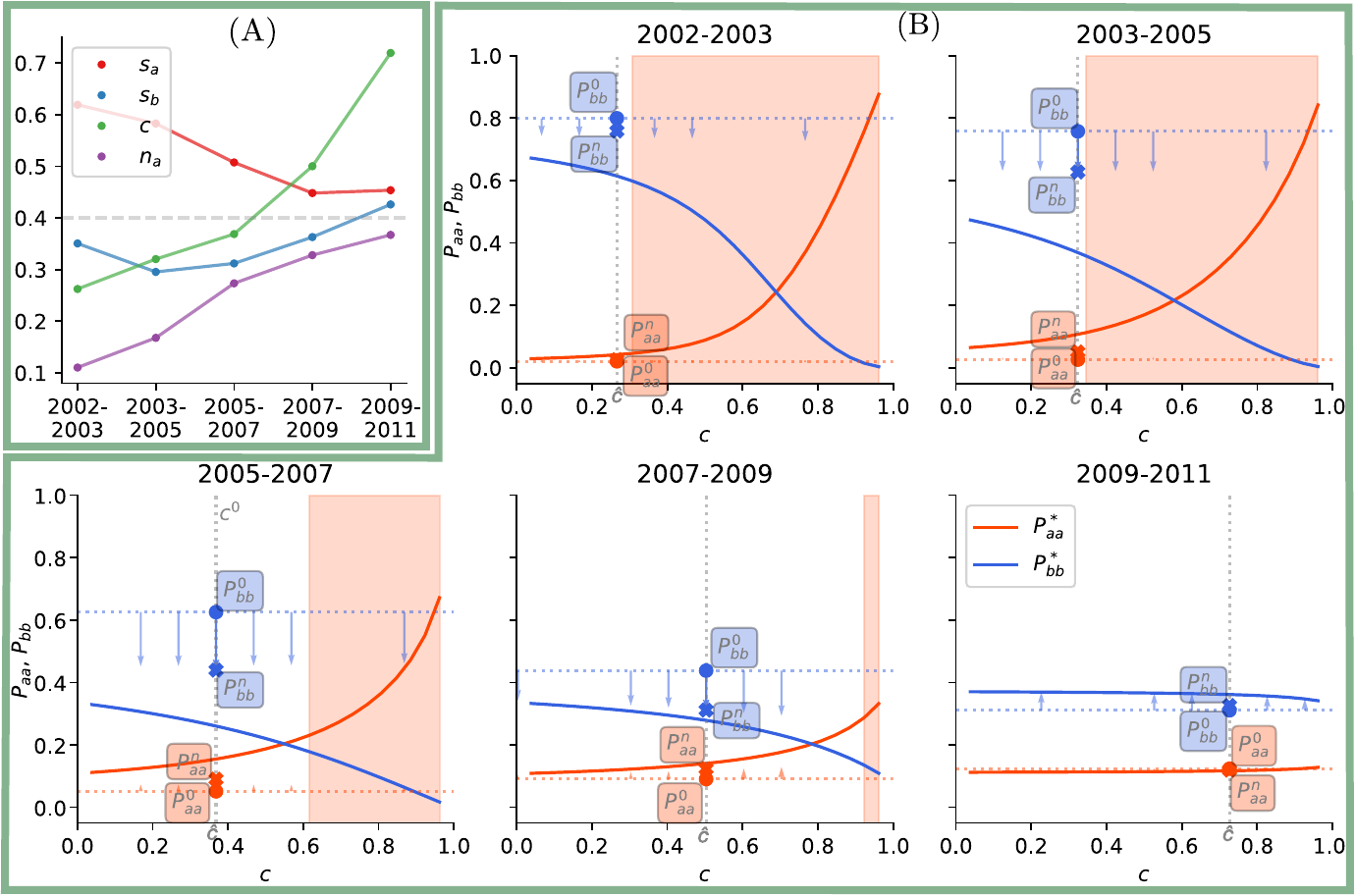}
\caption{\label{fig:fig5_boards}\small \textbf{Analysis of the \textit{Boards} dataset, including the effect of government policy on estimated parameters,  
and simulated policies affecting preferential attachment}. In this data, women are the minority group $a$ (red), and men the majority group $b$ (blue). (\textbf{A}) Evolution of our four fitted model parameters following five stages of an affirmative action government policy. (\textbf{B}) We sequentially display the evolution of the group mixing $P$ matrix at each time range (five panels with dates), and include the effect of varying 
preferential attachment values on the x-axis, where the line $c^0$ the estimated value. In the y-axis we depict the evolution of group mixing $P$, where large dots are the initial values $P^0$ and crosses the final values $P^n$.  
Fixed points of the group mixing matrix $P^*$ are depicted as solid lines, and shaded areas represent regions where the fixed points are in core-periphery with $a$ as the core. 
We compare  
the actual network evolution during the observation period (dots to crosses) 
to ``simulated policies'' (arrows) of increasing or decreasing $c$ given the initial conditions of the time range. Both simulated policies and empirical evolution tend towards the fixed points, with the empirical observations largely matching predicted trajectories.
}
\end{figure*}

Core-periphery is closely related to the concept of Old Boys' Club in social sciences, in which wealthy people with similar social and educational background help each other in business and restrict opportunities for outsiders \cite{lalanne2011old}. In social networks, people in the core are more likely to be connected to well-connected people, therefore they enjoy advantages that come with social capital such as access to novel ideas, job opportunities, and referrals \cite{zeltzer2020gender}. Here, we examine the use of our model in the \textit{Boards} network \cite{Seierstad2011}, which includes monthly data on boards of directors in Norway between 2002 and 2011. 
The groups are determined by two genders and a link reflects that two people belong to the same board. We contextualize three dates that reflect significant behavioural changes affecting our parameters: in 2004 women made up less than 12\% of the boards, triggering talks between the Norwegian government and companies to address such gender disparity. Since this issue was not addressed, in 2006 the government established a two-year period for reaching at least 40\% representation of each gender. On average, boards complied and this was the case after 2008~\cite{Seierstad2011}.

We examine the effect of policy interventions, namely to increase the minority size (fraction of women) per board, on the dynamics of core-periphery in the network. 
To do so, we analyze the \textit{Boards} dataset using the rewiring model (Fig. \ref{fig:fig5_boards}). Although there is only one direct policy, such an increase of the minority size per board entails several indirect effects in homophily and preferential attachment. We find that in the beginning women are much more homophilious than men (Fig. \ref{fig:fig5_boards}A, $s_a > s_b$), whereas at the end of the observation period, both genders largely converge in near-random mixing values ($s_a, s_b$ around 0.44). This is largely in line with the policy, as it encourages mixing by increasing the fraction of women per board, not at a global scale. However, we also find that the preferential attachment value $c$ largely increases over time. At the same time,  women still make up less than 40\% of all nodes (despite being, on average, at least 40\% of members in each board). This is on par with findings from \cite{Seierstad2011}, who observed that some women directors became ``prominent directors'' in the field, participating in several boards. Similar observations regarding positive homophilous tendencies among women has been also  shown in the case of Japanese board networks \cite{raddant2022interdependencies}.  

We also take the five temporal snapshots of the \textit{Boards} network and show how the data evolves (Fig. \ref{fig:fig5_boards}B).
The trajectories predicted by our model  
capture the changes in the data; by the last time interval, the behaviour seems to be near stationary. In the first two periods men capture between 80-60\% of connections, with a decreasing trend. 
In latter periods, dynamics show that $c$ has increased to $0.70$ and homophilies are near random mixing. At this stage, the effect of $c$ on the group mixing $P$ is small, likely changing the distribution of links within groups, not between them. 
Our model demonstrates that interventions that modify homophily and preferential attachment have a profound impact on the fixed points, leading to more favorable outcomes for women. Although the network structure is gradually approaching these fixed points, our model predicts that the transition to fully realize the potential of these new circumstances may be impeded by substantial structural inertia.
    
Our model allows us to examine the level of group mixing $P$ if there had been an alternative policy that would affect a different parameter, such as preferential attachment (we include full interventions in SI). This could happen, e.g., by restricting the number of boards that a single person can belong to, or by encouraging multiple board memberships. We track the evolution of the initial $P$ matrix from new $c$ values (red and blue arrows in Fig. \ref{fig:fig5_boards}). According to our model, increasing preferential attachment could result in women becoming a core at the fixed point if such a policy did not affect other parameters. In most cases, reducing $c$ would be disadvantageous to women as a group, as they would likely share an even smaller fraction of connections. This investigation can pinpoint scenarios in which boys' clubs (or elites in general) emerge in social networks, and examine ways to mitigate their effect. In the case of the \textit{Boards} network, the indirect increase of preferential attachment leads to some female directors becoming prominent, with an emergent ``elite'' not only determined by gender, but by participation on several boards \cite{Seierstad2011}.

\section*{Summary and Discussion}
        
Our findings show that core-periphery is an emergent effect of the interaction between two common network evolution patterns, assortative and preferential attachment. We focus on core-peripheries where groups are identifiable from extra-network characteristics, and where group behaviour is a driving mechanism in network evolution. We explore two minimal network evolution models that incorporate varying group sizes, preferential and assortative attachment, and basic mechanisms of real-world evolving networks: a growing model where nodes are added, and a rewiring model where links are redistributed. Our models disentangle assortative and preferential attachment by creating connections with a two-step mechanism. First, a focal node finds a candidate to connect to, selected with varying degrees of preferential attachment. Then, the focal node accepts or rejects the candidate via assortativity or choice homophily. Preferential attachment exposes focal nodes to ``popular", high-degree candidates irrespective of their group partitions, while assortativity determines whether the two nodes will actually connect.

These two mechanisms interact in a way that leads to core-periphery, even in the presence of two assortative groups (which would otherwise be largely disjoint communities). Once one of the groups starts gaining dominance in number of links, the focal node is mostly exposed to the dominant core, so links within that group are favoured by both mechanisms, making the periphery group lose even more connections relative to the core. This iterative process promotes cumulative effects that lead to a specific group gaining a dominant position in the network. This scenario can be triggered by several factors including the initial network configuration, one group being more assortative than the other, relative group sizes, or even random perturbations. 
    
We have found that group dominance differs qualitatively for growing and rewiring network dynamics. The growing model develops structures more akin to layered networks with high inter-group connectivity and diffuse cores, while the rewiring model produces cores connected mainly to themselves, and highly-disconnected peripheries. We largely attribute these differences to edge deletion when rewiring. Indeed, both models have the same mechanism for link creation, and differ when a neighbor detaches from the focal node. Over longer time scales, this means that links can effectively get ``transferred" from one group to another. Although real-world link formation and deletion might involve more elaborate mechanisms, the joint analysis of minimal growing and rewiring models sheds light on the various ways in which the interaction of assortative and preferential attachment leads to groups of nodes gaining dominant positions.

In empirical networks that display core-periphery, our model fitting shows that core nodes tend to be highly assortative, with the more distinct core-peripheries having a larger amount of preferential attachment. We have found evidence that some empirical core-periphery networks may be driven by an assortative core and a disassortative periphery, highlighting the importance of asymmetric group mixing. We have also observed that empirical data largely follows the core-periphery typology displayed by our theoretical models: hub-and-spoke networks with sharp cores for the rewiring dynamics, and layered networks for growing dynamics. 

Real-world networks probably display more diverse behaviours than our models suggest and are likely to be in different evolution stages. 
Importantly, we assume that group-level behaviour is a driving force in network evolution. However, group affiliations might not be static, binary, or defined as a single category. Correspondingly, when performing data analysis the available metadata might not fully align with the predetermined groups \cite{Peel2017}. 
As an example, in the \textit{Boards} network the homophilies tend towards random mixing but preferential attachment increases. Previous work has shown that at the latter stages a group of women became prominent in several boards. In this case, it would be interesting to assess whether different group partitions became relevant or whether only the increased preferential attachment was at play.
A scenario we don't cover is when emergent topological structures affect such node attributes. For example, in citation networks first-mover advantage \cite{Newman2009} may result in nodes gaining income or access to prestigious institutions. However, such scenarios would not explain, e.g., the role of elite education, race and gender in Old Boys' networks \cite{Reeves2017}, or the shift between centralized and decentralized social media dynamics with particular nodes at the core \cite{Bastos2018}. In many such cases, it is helpful to model group behaviour as a driver of social network evolution. Our results strongly suggest, however, that if highly-connected nodes become assortative, the combination of preferential and assortative attachment would further entrench their dominant position in the network. Other possible extensions to our model could include systematic late arrivals --a group of people arrives at a latter stage to the network--, as a way of assessing mechanisms for the social inclusion of minorities. 

It is possible that networked systems may not evolve purely through growing or rewiring mechanisms, and that more intricate versions of the link deletion mechanism exist beyond those included in our model.
Thus, we mostly avoided linking evolution mechanisms to mean-field predictions (we do so only for the \textit{Boards} dataset, where rewiring conditions resemble those of our models). Instead, we have compared the effect of removing preferential attachment on all networks. We find that preferential attachment amplifies existing disparities in assortative mixing,
increasing the share of links within the core for the rewiring model, and inter-group links for the growing model.

An example of how our model can potentially be used for real-world applications comes out of the analysis of the \textit{Boards} dataset, a historical scenario with a clear policy intervention that directly affected minority group sizes.
While the two gender groups became more likely to connect to each other (minimizing homophily disparities), the female group remained a nominal minority at a global level. 
Our model serves as an analysis tool for designing and understanding alternative interventions. Policies that decrease preferential attachment (by, e.g., limiting the number of boards a person can belong to within a time period) could have largely decreased the fraction of links within the minority.

Our findings demonstrate that the combined effects of preferential and assortative attachment can provide a compelling explanation for the emergence of core-periphery structures in networks. Furthermore, our models reveal that the different core-periphery typologies outlined in the literature \cite{Gallagher2021} can arise organically as a result of network evolution governed by these simple mechanisms. By understanding core-periphery structures as natural outcomes of network evolution, we can begin to extract valuable insights about how these networks were formed and developed.

\section*{Appendix}

\subsection*{Empirical data analysis}
    
We analyze datasets of two growing and three rewiring systems. For each case, multiple networks can be built. 
We systematically choose networks that display evidence of (i) core-periphery with groups determined by metadata and (ii) preferential attachment. 
We reject networks where the partitions induced by metadata do not display core-periphery, as they fall outside the scope of our study. However, it is possible that some topological core-peripheries do not align with the available metadata \cite{Peel2017}, whether because there are different mechanisms at play, or because the metadata is insufficient.

For (i), we construct networks for each dataset. We do not use Borgatti and Everett's correlation measure $r_g$, since it depends on the overall network density, making comparisons between networks difficult. Instead, we use a two-step approach based on group densities at their final stage. 
First, we identify whether a group $g$ is density-dominant in the sense that $\rho_g > \max\{ \rho_{g\hat{g}}, \rho_{\hat{g}} \}$ for $\hat{g}$ the other group. Second, we characterize the relationship to the periphery group using $\Omega_{g}$. For the growing network we choose networks where $\Omega_{g} > -0.1$, whereas for the rewiring scenario the criteria is $\Omega_{g} > 0.1$. The different cutoff values correspond to differences in the networks, since the growing models tend to concentrate more inter-group values in the datasets.
    
For (ii), given the ML method described in the main text, we test for evidence that the model with preferential attachment improves the description of the data over the model that only includes homophily estimates. To do so, we perform a likelihood ratio test for $c$, retaining networks that strongly reject $c=0$. 
    
\subsection*{Considered datasets}
    
We use five temporal datasets from which we analyze several core-periphery pairs. For the \textit{Airport}, \textit{ASP} and \textit{Cit-geo} datasets, many two-group sub-networks can be reconstructed, in which case we systematically analyze all possible sub-networks.
    
For the rewiring model we used a $\Delta t$ sliding window approach, where we construct a network using data for $\Delta t$. As we advance in time, we delete the links and nodes that are not within $\Delta t$ of the most recent item. The datasets for the rewiring model are:
    
\begin{enumerate}
\item \textit{Boards}: Boards of directors in Norway during 2002-2011, with monthly aggregation where groups are women ($a$) and men ($b$), with the gender determined either by analyzing the names or researching the individuals for ambiguous cases \cite{Seierstad2011}. We use this dataset for a detailed analysis and use known policy details to construct the networks, without filtering out by core-peripheriness. We use $\Delta t = 1$ year.
\item \textit{Twitter}: dataset from social media platform \textit{X} in Finland, collected between February 2020 and January 2021, when the platform was known as Twitter. The data collection included hashtags related to climate change discussions. Group $a$ are politicians from all political parties as well as political candidates from the previous election, while group $b$ are all other actors in the network. We construct several datasets with $\Delta t = 0.5, 1, 3, 6$ months, with ``retweet'' actions on the platform. Twitter is now known as X, we use the former name as it was the platform used during the data collection period. 
\item \textit{Airport}: Passenger use of airport network in the United States from the Airline Origin and Destination Survey (DB1B) from the United States Department of Transportation~\cite{airportdb1b}, with our dataset spanning the years 1993-1995. Nodes represent airports and links are flights between them. For the groups, we used nine regional divisions defined by the United States Census Bureau, we tested all pairwise interactions. We constructed networks on 2 year-periods, with $\Delta t = 1$ year. 
\end{enumerate}
    
For the growing model we construct networks by adding nodes and their associated links as they appear in the data. To account for parameter changes, we divide the data into five observation periods. Starting from the second period, we use all preceding data as initial conditions. The datasets used for the growing model are:
    
\begin{enumerate}
\item \textit{APS}: Citation networks for the American Physical Society (APS) journal of physics, where groups are different pre-defined sub-fields of physics. We use the version of the data obtained in 2015 \cite{kong2022influence}.
\item \textit{Cit-geo}: Citation network from publications in the Clarivate Analytics Web of Science database, where we focus on regional interactions during the period 1900-2010. We define groups using different geographical and political aggregations, including USA, Western Europe, Eastern Europe, Latin America, among others. 
\end{enumerate}
    
\subsection{Acknowledgments}
GI acknowledges support from AFOSR (Grant No. FA8655-20-1-7020), project EU H2020 Humane AI-net (Grant No. 952026), and CHIST-ERA project SAI (Grant No. FWF I 5205-N).
MK acknowledges support from Academy of Finland grant numbers 349366, 353799, and 352561. FK was partly supported by supported by the EU Horizon Europe project MAMMOth (Grant Agreement 101070285).
We acknowledge the computational resources provided by the Aalto Science--IT project. We used data from the Science Citation Index Expanded,
Social Science Citation Index and Arts \& Humanities Citation Index, prepared by Thomson Reuters, Philadelphia,
Pennsylvania, USA, Copyright Thomson Reuters, 2013.

\subsection*{Author contributions}
JUC, FK, GI, and MK conceived, designed, and developed the study. JUC implemented and analyzed the models and empirical data studies. JUC wrote the initial manuscript, and all authors contributed to writing the paper.

\subsection*{Data and code availability}
The code used in this paper is available at 
\url{https://github.com/javurena7/core-periphery-emergence/}. The data for \textit{Boards} is available on \cite{Seierstad2011}, \textit{Airport} on \cite{airportdb1b} and \textit{APS} on \cite{kong2022influence}. \textit{Cit-geo} is subject to a non-disclosure agreement and belongs to Clarivate Analytics Web of Science database. Non-public data from \textit{Twitter} (now known as \textit{X}) is available from the authors upon reasonable request.

\bibliographystyle{science}
\bibliography{prx}

\end{document}